\documentclass{acmtrans2m}

\usepackage{psfig} 

\acmYear{07} 
\acmMonth{January} 
\title{Recruitment, Preparation, Retention: A case study of computing
culture at the University of Illinois at Urbana-Champaign.}

\author{TANYA L. CRENSHAW and ERIN WOLF CHAMBERS \\
       University of Illinois at Urbana-Champaign \\[2ex]
       HEATHER METCALF \\
       University of Arizona, Tucson, Arizona \\[2ex]
       UMESH THAKKAR \\ 
       University of Illinois at Urbana-Champaign 
}

\begin{abstract}
Computer science is seeing a decline in enrollment at all levels of
education, including undergraduate and graduate study. This paper
reports on the results of a study conducted at the University of
Illinois at Urbana-Champaign which evaluated students attitudes
regarding three areas which can contribute to improved enrollment in
the Department of Computer Science: Recruitment, preparation and
retention.  The results of our study saw two themes.  First, the
department's tight research focus appears to draw significant
attention from other activities | such as teaching, service, and other
community-building activities | that are necessary for a department's
excellence. Yet, as demonstrated by our second theme, one partial
solution is to better promote such activities already employed by the
department to its students and faculty.  Based on our results, we make
recommendations for improvements and enhancements based on the current
state of practice at peer institutions.

\end{abstract}

\category{}{}{} \terms{} \keywords{}

\begin{document}

\begin{bottomstuff}
{\bf Contact author}: Tanya L. Crenshaw at
\emph{tcrensha@uiuc.edu}. \\ This work is supported in part by the
Department of Computer Science at the University of Illinois at
Urbana-Champaign and the National Center for Supercomputing
Applications.
\end{bottomstuff}
\maketitle

\section{Introduction}
The field of Computer Science is experiencing a downward trend in
enrollment at the university level \cite{Veg05}.  The Department of
Computer Science at the University of Illinois of Urbana-Champaign is
seeing a similar decline.  The authors conducted a study from January
to July 2006 which sought to determine through student surveys and
interviews how the department can increase these numbers.  The study
examined three specific areas, recruitment, preparation, and
retention. Seeing these three areas through students' eyes, we
pinpoint concrete recommendations for improvement for attracting
students to computer science, preparing them for industry or research,
and keeping them in the field.  This paper summarizes the findings of
the study and lists a set of recommendations for improvement at both
the undergraduate and graduate levels.

The paper is organized as follows.  We begin by describing the
organization and programs available in the Department of Computer
Science.  Section~\ref{s:rel} summarizes the related work.  In
Section~\ref{s:meth}, we describe the methodology used in the
departmental study.  Section~\ref{s:result} summarizes our results and
Section~\ref{s:rec} lists our recommendations.

\section{Department of Computer Science}
The Department of Computer Science at the University of Illinois is
frequently listed as one of the top programs in the United States, at
both the graduate and the undergraduate level \cite{USNews06}
\cite{USNews07}.  To demonstrate the current structural context of the
department, we describe the organizations supported by the department
and the undergraduate and graduate programs of study.

\subsection{Computer Science Organizations at UIUC} \label{ss:org}

The Department of Computer Science at UIUC supports a number of
organizations which provide academic, research, and social support and
outreach opportunities for students.  Throughout this document, these
programs are routinely mentioned by the authors and in the participant
data.  Although other organizations exist, the following briefly
describes only those discussed by our participants.

\begin{itemize}
\item {\bf CSGSO}: The Computer Science Graduate Student Organization,
or CSGSO, is a departmental organization whose goal is to improve
graduate life.  They host a weekly event, called Friday Extravaganza
(FE), which provides an informal event for graduate students to
socialize, as well as network with various industry representatives
whose companies sometimes sponsor the FE.  The CSGSO also sponsors
seminars on research and graduate life, although specific topics
differ annually.  Every graduate student is a member of CSGSO by
virtue of being in the department.

\item {\bf ACM}: The department hosts a student chapter of the
Association for Computing Machinery, or ACM, which currently has 350
members.  The ACM is open to both undergraduate and graduate students,
but is primarily populated by undergrads.  They organize weekly
meetings of special interest groups, or SIGS, on computing related
topics.  There are also general meetings to welcome new members and
inform current members of upcoming events, as well as social
events. ACM also sponsors a yearly conference, {\em
Reflections $|$ Projections}, which attracts corporate sponsors, academic
and industry speakers, and features a programming competition.  The
conference is attended by students from multiple universities.

\item {\bf WCS}: Women in Computer Science, or WCS, is an organization
which works towards recruitment and retention of women at all levels
of computer science.  WCS hosts meetings and social events for both
undergraduate and graduate students.  In addition, WCS sponsors
several outreach and recruitment programs, the largest of which is the
ChicTech program.  ChicTech organizes teams of students, predominantly
undergraduates, to speak at high schools about computer science and UIUC.  
ChicTech also hosts a competition for high school girls, who pick an
organization and design a software program or website for that
organization.  The teams then visit for a weekend at UIUC to
participate in activities and present their project.

\item {\bf !bang}: !bang is a computer science organization whose goal
is to foster more social activities within the Department of Computer
Science.  They generally host 2-4 events per year, usually also
sponsored by ACM and WCS.  The events are large-scale departmental
events for students and professors.  In recent years, !bang has hosted
bi-annual !casino, which most recently was sponsored by Google.
Professors from the Department of Computer Science volunteered as
dealers for blackjack, poker, and roulette tables.  The event allows
for interactions between professors, undergraduates, and graduate
students.

\item {\bf WIE}: Women in Engineering, or WIE, is an organization
sponsored by the College of Engineering whose goal is to promote
recruitment and retention of women in all departments within the
college.  WIE hosts the Freshman Camp, a weekend retreat at the
beginning of the year for all incoming freshmen women.  Current CS
students, organized through WCS, attend this camp to meet the freshmen
and offer encouragement and advice.  WIE also organizes a two week
program for middle school girls called, G.A.M.E.S., or Girls'
Adventures in Mathematics, Engineering, and Science.  There are two
specialized camps, including one focused on Computer Science.  Many
students and professors in the Department of Computer Science assist
as camp counselors and mentors.

\end{itemize}

\subsection{Undergraduate program}

The undergraduate program in the Computer Science department at UIUC
is a four year program.  All undergraduates take the same core classes
in their freshman and sophomore years, including programming, data
structures, discrete math, theory of computation, architecture, and
computer ethics. For upper level students, there are currently two
curricula.

Most current juniors and seniors are on the older track, where
students take programming languages, operating systems, analysis of
algorithms, an advanced architecture course, and two additional
400-level courses of their choice.  The older track also has an
application sequence, a set of approved courses in another discipline
to which computer science can be applied in a meaningful way.  Typical
application sequences can be in engineering, mathematics, psychology,
music, or business, but students may also design their own
application sequence based on their individual interests.

For those students entering after Fall 2005, the course requirements
are quite different.  The application sequence is no longer a part of
the program; instead, students choose a technical track emphasizing a
particular area of CS in which they are interested.  The three tracks -
computer science, scientific computing, and mathematics - have
different advanced course requirement appropriate to their
specialization.  However, since this is such a new program, no
students surveyed have progressed far enough on this track to comment
on these upper level distinctions.

In addition to majoring in Computer Science, undergraduates can obtain
degrees from the College of Liberal Arts and Sciences as a Bachelor of
Science in Math and Computer Science, or a Bachelor of Science in
Statistics and Computer Science.  Moreover, there are other programs
available, including a minor in Computer Science that is available
campus-wide (except to students majoring in Computer Engineering), as
well as the Software Engineering Certificate.  

\subsection{Graduate program}

The graduate program has multiple programs, both for a Ph.D. and a
Master's degree.  Undergraduates at UIUC can opt for a fifth-year
Master's degree, taking 16 hours of additional course work and writing 
a thesis.  Enrolled graduate students can obtain a Master of Science 
degree, a research-oriented degree requiring 28 hours of graduate 
course work and a thesis, or a Master of Computer Science, requiring 
only 36 hours of graduate course work.  For either degree, there is a 
distribution course work requirement in which students must take a course 
in the areas of Software, Architecture, and Theory and obtain a grade 
of B- or higher.

For a Ph.D., students must take a total of 48 hours of graduate
course work and 32 hours of thesis work to obtain the degree.  These
requirements are reduced for students entering with a Master's degree.
The Ph.D. course work requirements differ for students entering before
and after Fall 2005.  Before Fall 2005, students have to take one
course in the areas of Programming Languages, Operating Systems, and
Theory and any other area of their choosing.  Students have to achieve
a grade of at least A- in three of the courses, and B- in the fourth.
After Fall 2005, students have to take one course from Theoretical
Computer Science and Formal Methods, and one course from Systems and
Architecture.  Students must also take two courses in their research
area, at least one of which must be a 500-level course.  For 400-level
courses, students must obtain a B+ or better.  For 500-level courses,
students must obtain an A- or better.  UIUC also offers a
Ph.D./M.D. degree, but no students pursuing this degree were surveyed
or interviewed.

The distribution requirement for Master's and the foundational
requirement and research course work requirement for Ph.D. will be
loosely referred to in this document as ``core course work.''

\section{Related Work} \label{s:rel}

The vast majority of research on the decreasing enrollment throughout
computer science education repeatedly focuses on two potential
"solution" areas: recruitment and retention, which involve getting
more students to join and keeping those who have joined \cite{Cohoon}
\cite{Margolis} \cite{Blum}.  Recruitment entails attracting K-12
students to computer science and often involves outreach efforts which
attempt to make computer science look "cool," exciting, useful, and
rewarding.  In addition, when recruiting underrepresented students in
particular, researchers often recommend opening up admissions criteria
without lowering standards, welcoming reentry students, and providing
opportunities to bridge educational gaps that students might have
between their previous education and the entry-level courses at the
university \cite{Margolis} \cite{Cuny}.

The goal of retention is to keep the students and computer scientists
already in the field.  Retention efforts involve various support
structures for the existing students, particularly those who are
underrepresented and are more likely to leave. There are several ways
in which these support structures can be fostered; providing role
models via good teaching, advising, mentoring, and outreach
\cite{Ragins} \cite{Camp} \cite{Blum}.  With teaching, students can
experience the enthusiasm of the subject through their professor or
instructor.  Positive advising relationships can help foster a
graduate student's self-confidence and research success.  For
undergraduates in computer science, advising not only helps the
student select courses and fulfill requirements, but also frequently
determines the level of interest in pursuing various career options or
even a graduate degree.  For graduates, positive advising
relationships can result in more productive and happier students.
With mentoring, students can reach out to their more experienced
colleagues and faculty for support.  Mentoring is not just about moral
support, but also has an impact on whether students finish a program,
get good advice, and feel happy about their education \cite{Jaschik}.
Finally, with outreach, students can use their expertise in computer
science to benefit others, thereby increasing their own
self-confidence and obtaining real-world examples of how their
computer science skills can be used.

A third, less researched area for solving the problem of decreasing
enrollment involves working to create a flexible culture of computing
which is open to diversity and allows for students and faculty to
define for themselves what it means to be a computer scientist
\cite{Margolis} \cite{Blum}.  Margolis explains that ``one of the aims
of higher education must be to provide students with a broad picture
of possibilities and to create an environment where alternate models
are valued and respected.''  One way in which a program can create such
a culture is to improve its gender balance.  A cross-national study of
male overrepresentation in Computer Science \cite{Charles} starkly
demonstrates the influence that even a nation's culture can have on a
woman's entrance into the field.  The degree of male
overrepresentation in the Czech Republic is three times that of
Turkey, the country with the most gender-integrated program.
Diversity in race, socioeconmonic status also contribute to the
environment that Margolis describes, and whether or not students join
and stay in the field.

As noted above, the abundance of literature focuses on recruitment,
retention, and the newer and broader idea of computing culture.  Much
of the related work focuses on either the ``problems'' of computer
science culture, or the ``solutions'' which make computer science look
cool, but lacks consideration of students' attitudes regarding the
problem areas of computer science enrollment.  This paper describes
another approach.  The study was designed to investigate and
understand the attitudes of undergraduate and graduate students in
three areas which are well understood to position students for
success; recruitment, preparation, and retention.  These areas include
activities such as departmental culture, teaching, advising, and
mentoring.  Once their attitudes were understood, our explicit goal
was to uncover actual, working solutions which members of the
department can cooperatively utilize to reverse the decline in
enrollment.

\section{Methodology} \label{s:meth}

The study took place in two phases from January to July of 2006.  In
the initial pilot study, four undergraduates and seven graduate
students were individually interviewed for a single hour. By no means
were the participants selected randomly.  Rather, for our pilot study,
participants were hand-picked by the researchers to get a breadth of
experiences from sophomore undergraduates having just started their
computer science courses, to sixth-year graduates about to complete
their Ph.D. work, from males to females, and from parents to
non-parents.  Among undergraduates, the students interviewed were
100\% women.  Among the graduate students, the study was approximately
57\% female and 43\% male.

\begin{figure}[h]
\centerline{\psfig{figure=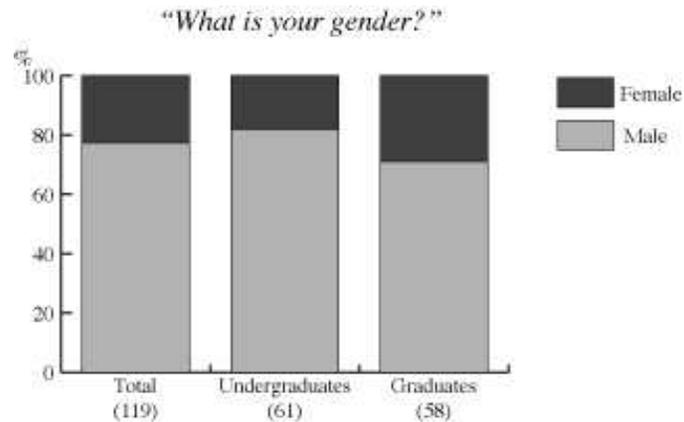,width=3.5in}}
\caption{Overall, in the second-phase, on-line survey, participants
were 17\% female and 83\% male.  Female participants were oversampled
by announcing the survey to WCS and female graduate students one week
before the survey was announced to the general student population of
the department. The gender ratio here is not an accurate reflection of
the department.%
    \label{f:demograph}}
\end{figure}  

The second phase consisted of an on-line survey developed as a result
of the pilot study, combined with nine interviews to supplement the
survey data.  Participants were recruited randomly through
departmental e-mail. Female participants were oversampled by
announcing the survey to departmental, female-only groups one week
before announcing it to the general student population.  For their
participation in the survey, students received a three dollar gift
certificate to a local coffee shop, and for an interview, students
received a five dollar gift certificate.

The participation in the survey is summarized in
Figure~\ref{f:demograph}.  A total of 119 students participated in the
survey, 61 undergraduate students, and 58 graduate students.  Overall,
the survey participants were 17\% female and 83\% male.  Because the
female population was oversampled, this ratio is not an accurate
depiction of the actual gender ratio in the Department of Computer
Science at UIUC.

All figures which appear throughout the remainder of this document are
taken solely from the data obtained through in the on-line study
results gathered in the second phase of the study.  Where appropriate,
quotations from the interviews are used to exemplify the themes of our
results.

\section{Results} \label{s:result}

The results of our study have been organized into three areas which
impact student success in computer science.  The first area is
{\bf recruitment}, the means by which participants were initially
attracted to and became members of the field of computer science.
Recruitment has an obvious impact on a student's success, for a
student cannot be successful in computer science without first
entering the field. We examine how students discovered computer
science, and how they discovered the University of Illinois as an
institution for study.

The second area, {\bf preparation}, describes the resources a student
needs to prepare for his vocation of choice in computer science.
These resources are course work, including the quality of teaching,
and early research opportunities, each of which has an impact on a
student's success.  For undergraduates especially, course work allows
students to experience the enthusiasm of a particular subject through
a professor or instructor.  In addition, a curriculum's design, can
impact the flexibility a student has to explore her own interests.
For undergraduates, early research opportunities allow a student to
experience a taste of academia, and encourage her to enter into
graduate school.  For graduates, early research opportunities allow
them to make a faster transition from absorbing knowledge to creating
it.  We examine students attitudes regarding their access to these
resources in the department.

The third area is {\bf retention.}  We define retention as a student's
feeling of membership both in the department and in the greater
computing community.  The methods by which this membership can be
cultivated are advising and mentoring relationships, as well as the
quality of a student's environment and work-life balance.   All
of these impact a students' success.  Advising can help a student to
select a career in which they can be productive and fulfilled, or to
make better and more significant progress on research.  Mentoring
gives the student access to a more senior member of the computing
community, one who can give good advice or lend moral support.
The environment, such as a classroom of peers, can further help or
possibly hinder this membership.  Finally, allowing students a healthy
work-life balance gives them time to participate in the outside
activities they value.  Again, we examine students' attitudes
regarding their access to positive advising and mentoring
relationships, as well as their environment and work-life balance.

\subsection{Recruitment}

Within the theme of recruitment, we investigate three areas: how
students discovered computer science, why graduate students in
particular selected UIUC as a school for their studies, and how
participants take part in recruitment.

\subsubsection{Discovery of Computer Science}

\begin{figure}
\centerline{\psfig{figure=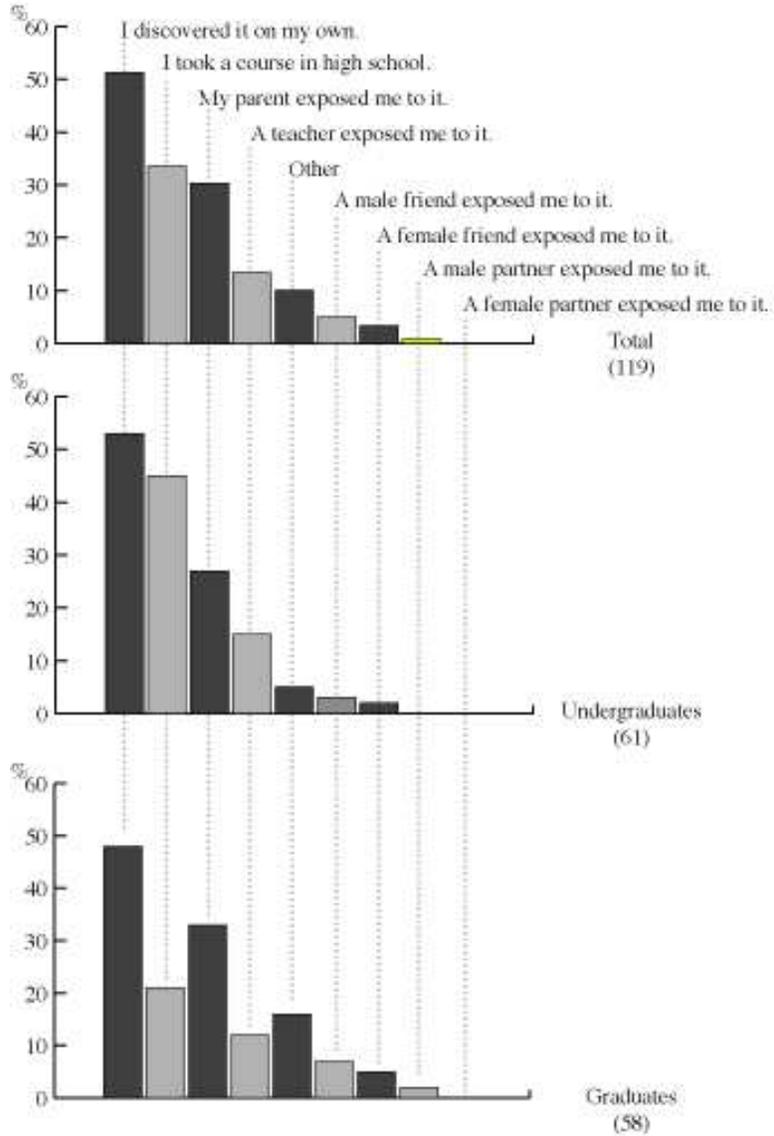,width=4in}}
\caption{Participants were asked how they were first exposed to
computer science.  The top three choices for both undergraduates and
graduates were self-discovery, high school course, and parents. More
undergraduates than graduates reported taking a course in high school,
and understandably so as more high schools have offered computer
science courses in recent times. In all cases ``Other'' referred to
being exposed to it via a brother, or via a previous job.%
    \label{f:discovery}}
\end{figure}

As summarized in Figure~\ref{f:discovery}, participants varied as to
how they were initially attracted to the subject of computer science.
The majority of the participants report discovering computer science
on their own. Many had parents who were computer scientists and were
attracted to the field at an early age.  Others report being exposed
to it by a teacher or friend.  One participant studied Information
Systems as an undergraduate, but changed to computer science so he
could join his wife at graduate school.  Responses for the ``Other''
category shown in Figure~\ref{f:discovery} included being exposed to
computer science by a brother or via a previous job.  It's striking to
note that no participant reported being exposed to computer science by
a female partner.

\subsubsection{Graduates' discovery of UIUC}


For the graduates, the method by which students selected UIUC for
graduate school is best described by the following interview
participants,

\newpage
\begin{center}
\emph{I got the list of--US News or whatever--and I got a
list of the top 25 $\ldots$ and UIUC was one of those.}
\end{center}

\begin{center}
\emph{UIUC was the highest ranked school I applied to, and I was lucky
to get into it. It was a real obvious choice for me.}
\end{center}

\begin{center}
\emph{As everyone else, I looked at US News [and World
Report] and I looked at what the rankings were.}
\end{center}

As seen in the above, the response for why students selected UIUC for
graduate school was generally ``It was the highest ranked place I got
into.''  One participant said of the prospective student visit, ``I
liked it a lot when I visited it.  I felt like I fit in with the
people, mostly the students.''  Another participant echoed this
statement and went on to report that it also helped that a particular
faculty member expressed excitement in her attending UIUC during the
prospective visit.

\subsubsection{Student attitudes and participation in recruitment}

To determine to what degree participants currently participated in
recruiting activities, which we termed ``outreach'' in the survey, we
began by asking participants if they had ever used their
expertise in computer science to benefit other people.  The examples
given in the survey question were helping friends with homework,
taking a prospective student to lunch, or volunteering at the local
Boys and Girls Club. Of the undergraduates surveyed, 79\% said that
they had.  When asked about details of their schedule, 18\% said they
spent 1 to 5 hours a week working on outreach activities.  Of the
graduate participants, 73\% replied that they had, and 18\% reported
spending 1 to 5 hours per week working on outreach activities.

All of the undergraduate interviewees understood outreach to mean
``exposing computer science to the outside world.''  All
undergraduates interviewed felt that outreach was very important, and
all but one also thought it was very important to the department.
Most of those interviewed had attended or helped with the Freshman
Camp, and thought that it was a helpful program.  All were aware of or
had assisted with G.A.M.E.S.; one had even first been recruited to
come to UIUC through that program.  Two participants had assisted with
recruiting trips to Illinois high schools to talk to juniors about
computer science at UIUC.  All of them said that they participated in
outreach in some fashion, whether through the WCS and WIE activities
already described or by helping students not majoring in computer
science with their programming projects.

During the graduate interviews, however, the terms ``outreach'' or
``recruitment'' were construed two ways.  First, as with the
undergraduates, ``outreach'' was understood as a means to expose
computer science to the outside world.  Second, it was viewed as
interactions within and between departments which result in a student
support system.

Graduate student interview participants were not very aware of the
existing outreach from the first category.  Some noted the ``G.A.M.E.S. 
Camp'' as one existing outreach program.  One participant said of her
participation in outreach, ``I would if I were asked.''
Another participant said simply, ``We don't do it,'' pointing to
either a lack of good advertising on the department's part or a lack
of attention on the participant's part.  Another simply stated, ``I would
like to see more active recruitment and retention of underrepresented
groups.''

In response to questions about the second type of outreach and
recruitment, graduate student interviewees also expressed a desire to
see more inter- and intra- departmental interactivity as a way to
support the students.  In this regard, one participant said,

\begin{center}
\emph{One of the things that surprised me when I got
here was how little interaction there is among students, especially
across research groups. It is a large department, and I had very
little idea what other students, especially students who had been in
the dept long enough to be doing research, were doing. I think I would
have benefited a lot from seeing the kinds of projects other students
were working on, and talking to them about how to go about getting
started with my own research.'}
\end{center}

Another echoed this statement with a wish for, ``Better introduction
to research for new students, faculty involvement in social
activities, more interdisciplinary events with the arts and
humanities.''  Yet another echoed it with the suggestion to, ``Promote
communication among senior and junior phd students.''  In fact, 31\%
of the graduate students pointed to a need for some kind of
improvement for faculty-student or student-student interaction.

Participants interviewed offered suggestions about how more
``outreach'' and ``recruitment'' could be incorporated into the
department.  One participant said the department should perform
outreach to local industry to help teach concepts such as ``rapid
prototyping which would also help students with networking.''  This
tied to his opinion that the department should be ``selling the
students more.  Other schools have a lot more networking.  We don't
teach our students how to network.''  Another participant, who after
her preliminary exam had attended only one conference, said this about
her advisor's networking for his students,

\begin{center}
\emph{I haven't seen any evidence that he has, so I assume that he
hasn't.  I guess that the typical thing that I see is when you go to a
conference and you take your student and you start introducing your
student to everyone, and that has never happened [to me].  If I was an
advisor, and I had any students, I would do this.  That one conference
that I went to, he wasn't there.  I listened to people's talks, but I
couldn't make any connections.}
\end{center}

The inability to make such connections in the larger community can
have serious effects on a student's ability to pursue a career in
academia.

Connecting students not only to professors, but also to each other can
help them to flourish.  One participant wished the department had more
group activities for the first-year graduate students to encourage
camaraderie, allowing students to ``outreach to each other.''  She
said, ``The CSGSO is not working adequately or properly'' and noted
that the Friday Extravaganzas were ``not a good way to meet new
people.''  She expected more activities would be offered by the
department for the first year students since ``there was such a nice
program for the prospective students.''

The study uncovered a definite disparity between the attitudes of
undergraduates and graduates regarding outreach.  This disparity seems
to be a direct result of the active undergraduate society chapters.
All but one of the undergraduates interviewed were active members of
WCS and ACM, and most participated in outreach through those
organizations. While these organizations are open to graduate
students, they mostly target undergraduates and, as we will discuss
later, most graduate students from this study expressed difficulty in
finding space within their research, work, and course schedules to
participate in such organizations.  Though graduate students seemed
largely unhappy with the activities organized by CSGSO, most of the
undergraduates enjoyed the societies in which they participate, and two
even echoed the statement that, ``without WCS, I probably would not
still be in CS.  After all, you can't go out and promote CS to high
school students and then drop out of the program yourself.''

\subsection{Preparation}

To understand how well the department is preparing its students, we
examined students' future career interests, as well as their attitudes
regarding their access to the important resources of teaching,
course work, and early research opportunities.

\subsubsection{Future Plans}

\begin{figure}
\centerline{\psfig{figure=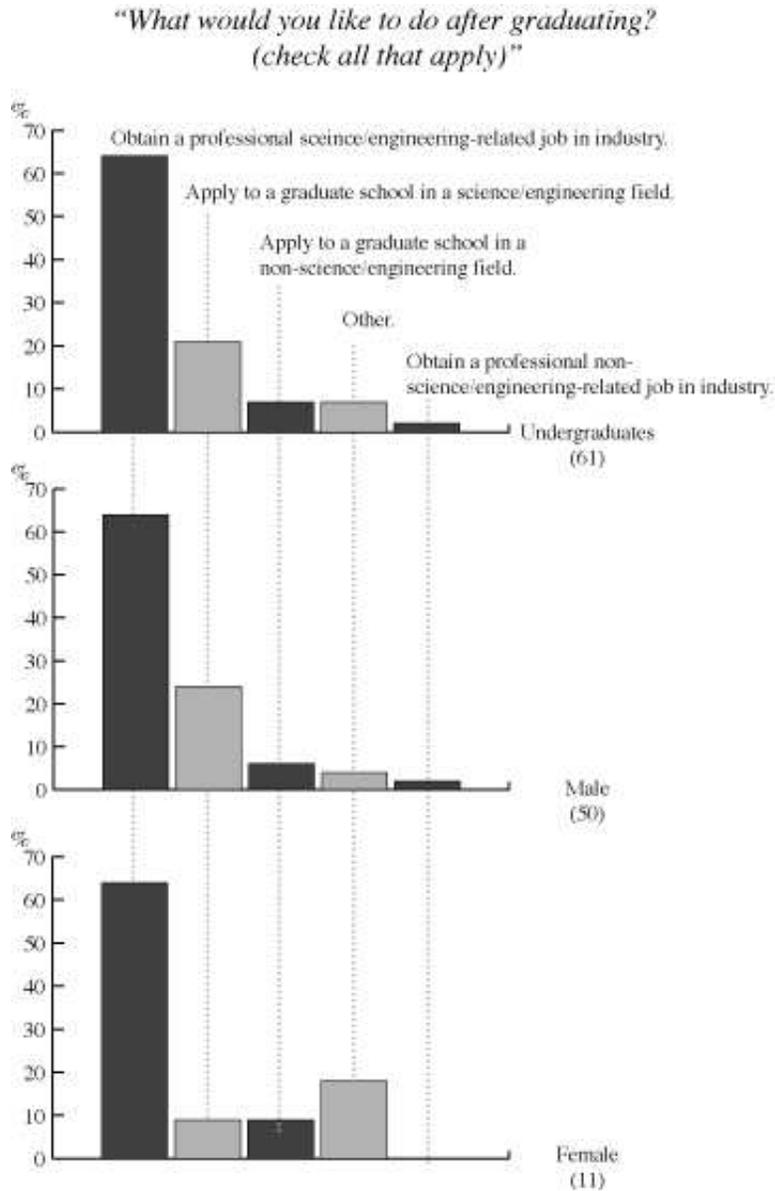,width=4in}}
\caption{Undergraduate participants were asked to select multiple
choices from a set of possible career options they would consider
after graduation.  The top option for all participants was ``Obtains a
science/engineering related job in industry.''  Significantly more
male than female undergraduates reported considering applying to
graduate school in a science/engineering-related field.  Female
undergraduates second most popular choice was ``Other''; choices
included going to law school, and starting a family or a business.
Both male and female undergraduates included ``I have no idea'' as an
``Other'' option. %
    \label{f:future_ugrad}}
\end{figure}

To determine how well UIUC prepares its students for their future
vocations, we examined students' future plans.  In the on-line survey,
students were asked what kind of job they were considering after
graduation.  Participants selected any job from the list that they
were considering, thereby allowing for multiple answers.
Undergraduates and graduates were offered different sets of possible
career choices.  Figure~\ref{f:future_ugrad} summarizes the future
plans of the undergraduate participants. The top choice of the
undergraduate participants was ``Obtain a science/engineering related
job in industry.''  Significantly more male than female undergraduates
reported considering applying to graduate school in a
science/engineering-related field, with 24\% male and 9\% female
participants reporting they would consider this option.  Female
undergraduates' second most popular choice was ``Other''; choices
included going to law school, and starting a family or a business.
Both male and female undergraduates included ``I have no idea'' as an
``Other'' option.

\begin{figure}
\centerline{\psfig{figure=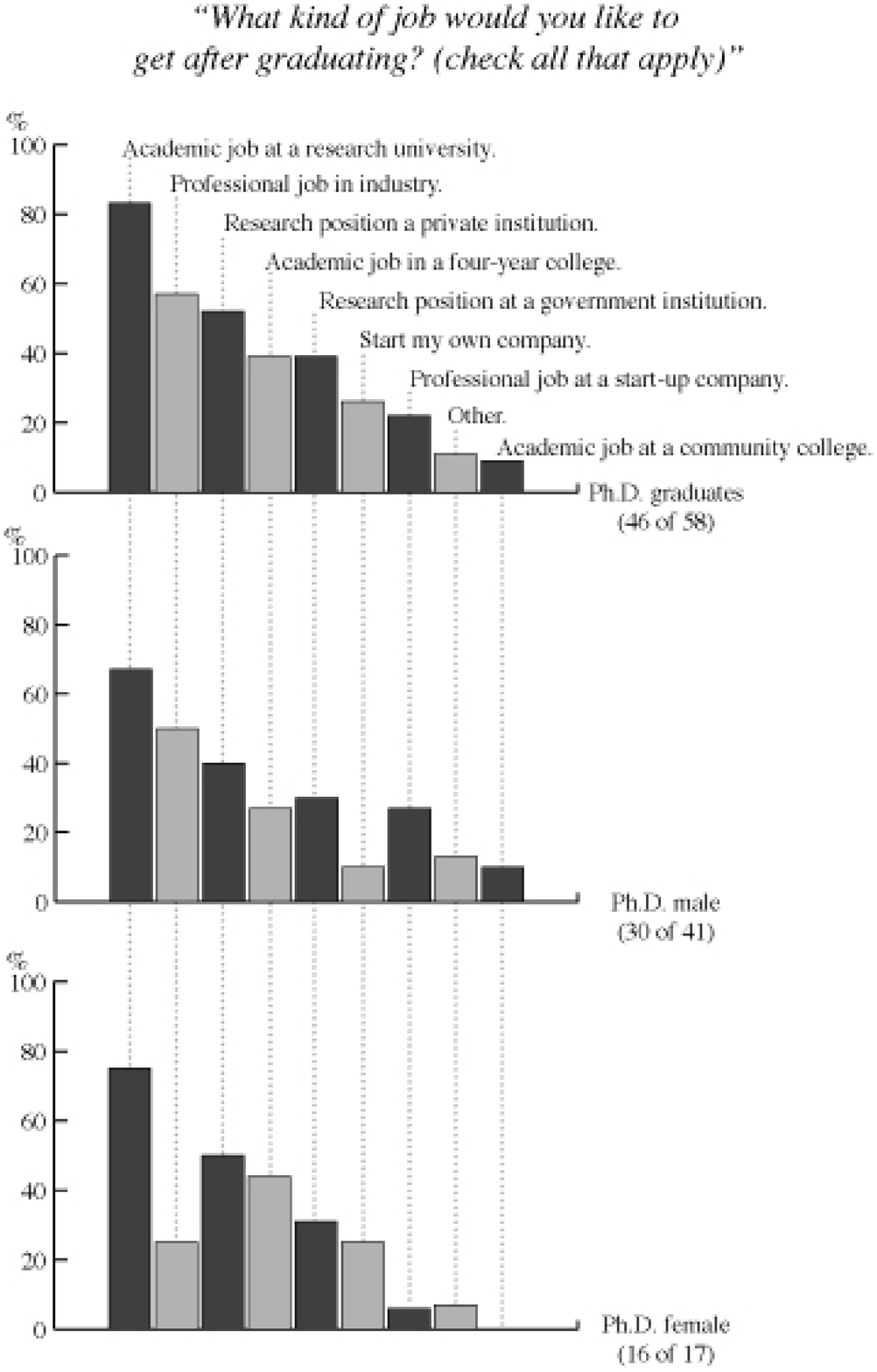,width=4in}}
\caption{Graduate participants in the Ph.D. program were asked to
select from a set of possible career options they would consider after
graduation.  Of the 46 of 58 graduate students who were
Ph.D. candidates, an academic job at a research university was the top
choice.  Female participants predominantly chose this option, while
male choices were more distributed across the options.
    \label{f:future_grad}}
\end{figure}

Figure~\ref{f:future_grad} summarizes the future plans of the graduate
participants in the Ph.D. program.  For the total group of
participants, an academic job at a research university was the top
choice.  Female participants predominantly chose this option, with
75\% of females reporting that they would consider this choice. Male
choices were more distributed across the options.  The second most
popular choice for males was ``A professional job in industry'' while
the second most popular choice for females was ``Research position in
a private institution.''

\newpage
\subsubsection{Teaching}

\begin{figure}[h]
\centerline{\psfig{figure=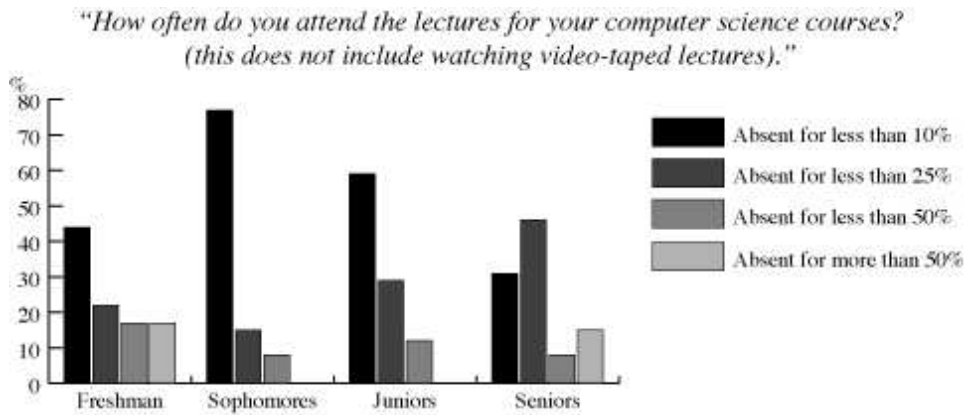,width=5in}}
\caption{Undergraduates were asked to report how often they attend
their lectures.  Sophomores attended their lectures the most and often
remarked on the quality of teaching in their CS173, \emph{Discrete
Structures}, and CS225, \emph{Data Structures}, courses.%
    \label{f:attendance}}
\end{figure}

To determine to what extent teaching prepared the students, we asked
undergraduates and graduates about their experiences with teaching in
the department.  The undergraduate students were asked, ``Do you think
that the Department of Computer Science values excellent teaching.''
Of the participants, 65\% replied yes, with more freshman and
sophomores replying in the affirmative than juniors and seniors.
Undergraduate students also reported on their attendance in lectures
as summarized in Figure~\ref{f:attendance}.

Reasons for not attending lectures included, 

\begin{itemize}
\item ``I do not have time to attend my classes.''  (25\%)
\item ``I do not feel that the lectures help me to learn.'' (66\%) 
\item ``I do not like the teaching style of the lecturers.'' (49\%)
\item ``I prefer to watch lecture videos.''  (8\%)
\end{itemize}

Many students also selected an ``Other'' reason for not attending
class, 13\% of which cited an inability to wake up in time for class,
or a preference of sleep over lectures.

For the undergraduates, the teaching quality in the department was one
of the most heavily discussed topics during both the interviews and
the open-ended questions on the on-line survey.  More than half of the
suggestions for how to improve the department were related to their
courses. In particular, better-trained teaching assistants, more
interactive lectures, and more collaborative assignments were the main
areas of improvement identified by the undergraduates. Representative
comments include:

\begin{center}
\emph{I know that the department is trying to keep a high ranking, but
sometimes I feel that the ranking is the only thing important to the
department. The department needs to focus more on the students than on
the rankings.}
\end{center}


\begin{center}
\emph{Lots of people feel/act like an island.  It would be better if
the department cultured more group activities in the classroom.}
\end{center}


\begin{center}
\emph{Fix the teaching, annotate all powerpoints during class, try to move
away from powerpoints and towards writing on a tablet/touchscreen so
that the students can see the thought process that's going on into the
explanations as the professors are doing it. Likewise, for code
examples type them up on the fly if possible.}
\end{center}

\begin{center}
\emph{I would like to see T.A.s with more instruction on how to teach
a class.  The first course a student takes in computer scientist is
the most critical, because it is that course that will make a student
decide whether to stay in the department.  These courses are often
considered 'weed-out' courses that get rid of the students who are
unsuited to computer science.  However, in my experience, they weed
out some of the most creative students, simply because the way the
course material is presented makes the students feel inadiquet [sic]
and dumb.}
\end{center}

\begin{figure}[h]
\centerline{\psfig{figure=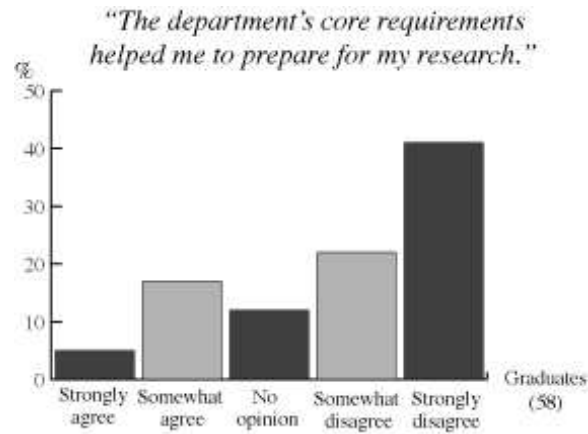,width=3in}}
\caption{Participants were asked how much they agreed with the
statement, ``The department's core requirements helped me to prepare
for my research.''  Participants predominantly strongly disagreed.%
   \label{f:core}}
\end{figure}


Graduate students, whose requirements focus around taking certain core
courses early in the program, were especially critical of the
department's attitude towards teaching.  One participant said,

\begin{center}
\emph{I think this department has some huge problems that are going to
catch up to them.  The first is the underlying culture in the faculty
that teaching and education are not high priority or important to the
image of the school.  The second is the high number of young
professors which contributes to the first problem because they are so
focused on getting tenure.
}
\end{center}

Another participant said she expected UIUC to have good faculty and
courses, but got ``badly taught, unorganized courses in which the
professors didn't prepare.'' She wished that the department ``enforced
better instruction'' but admitted that not much could be done if a
``professor were too famous.''  Another said that the 400-level
courses are ``disorganized, have no textbooks, and there is no
feedback.''  Another said, ``I got a grade in the class, but my
teacher didn't actually give back any homework grades.''  When asked
about her future plans, one participant said, ``I want to go to a
university where teaching and research is supported, not like UIUC.''

Most graduate participants were particularly unhappy with the core
course work as reflected in Figure~\ref{f:core}, with 41\% reporting
that they ``strongly disagreed'' with the statement that the ``core
requirements helped me to prepare for my research.''  During the
interviews, graduate participants reported a frustration with the
delicate balancing act between research and course work in their first
two years.  Many participants were particularly vehement regarding the
demands of the core requirements, expressing frustration with being
worried about getting high grades when they would rather be conducting
research.  The combination of demands of course work,
teaching-assistantships, and the core grade requirements were pointed
to as reasons for graduate students not having publications until
their third or fourth years.

\begin{center}
\emph{What you are expected to take is asinine.  I think it's an
embarrassment to the department that I can get to my qualifying exam
without ever having seen [a fundamental concept in my
area]\footnote{Topic omitted to protect identity.}.  The core
curriculum is especially frustrating for students because the courses
are poorly taught.}
\end{center}

\begin{center}
\emph{They make you take two full years of classes and then [the
faculty] get pissed off at you when you don't have a publication your
third year?}
\end{center}

\begin{center}
\emph{Get rid of the core classes altogether or make a significant
reduction.  I could spend my time preparing for research in my area
rather than wasting my time taking classes that are in NO WAY related
to my research.  If I'm lacking in a needed area, let the qual
comittee [sic] decide.
Let me explore the research areas on my own rather than forcing me to
take core classes that professors don't care to teach and students
don't care to take.}
\end{center}

Participants suggested that the core course work be made more flexible.
One participant suggested that the core be changed to a series of
specialized tracks based on the students' area of interest.  For
example, ``Someone in architecture should take VLSI, architecture, and
compilers.''  Another participant thought it would be worthwhile to
``take a course not in your area'' but thought it was ``dumb to have
to take a systems course.''

\subsection{Retention}

We define retention as a student's feeling of membership in the
department and the greater computer science community. We examined
whether or not participants felt that they were a member of the
community via their attitudes regarding their access to positive
advising and mentoring relationships, as well as their environment and
work-life balance.

\subsubsection{Advising}

Undergraduates in the CS department are assigned a faculty advisor
when they first enter the program. In the past, meetings with this
advisor were not required, but were just encouraged.  Today, freshman
and juniors are required to to meet with their advisor to discuss
topics ranging from course work to potential career tracks. There is
also an undergraduate advisor available in the main office to meet
with students and address their advising needs.

In the second phase of the survey, only 2 undergraduate participants,
or 3\%, said that their advisor also served as a mentor to them. Most
students interviewed did not meet with their advisor, but instead
relied on peer advising. In the pilot study interviews, students
were asked if and why they had considered leaving the department. Half
of the undergraduates interviewed reported that they did not get
enough support from their advisor or the faculty they knew. One
participant reported that she didn't feel she had anything to discuss
with her advisor; that he didn't know enough about the classes to get
any help planning her curriculum.  A single participant interviewed
reported a positive interaction with faculty; it is notable that this
participant was also the only undergraduate with any research
experience.

\begin{figure}[h]
\centerline{\psfig{figure=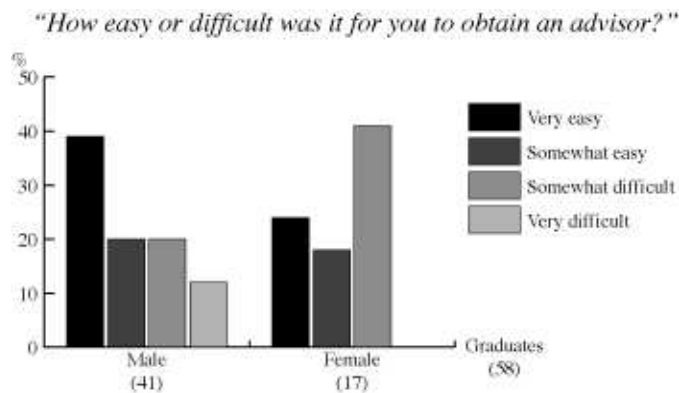,width=3.5in}}
\caption{Graduate participants were asked to rate the ease or
difficulty with which they obtained an advisor.  Of the participants,
12\% of males had a very difficult time, while 41\% of female
participants had a somewhat difficult time.
    \label{f:ease}}
\end{figure}

For graduate students, an advisor plays quite a different role than
for undergraduates.  Graduate students must seek out their own
research advisors by their third semester and work with them until the
end of their degree.  Simply obtaining an advisor can be a challenge,
as is reflected in Figure~\ref{f:ease}.  Of the graduate participants,
12\% of males had a ``very difficult'' time obtaining an advisor,
while 41\% of the female participants reported a ``somewhat
difficult'' time.

During the interviews, many graduate participants echoed the feeling
that ``there are more students than there are advisors.''  One
participant said that she spoke to ``five or six professors before
getting an advisor.''  Another participant reported she was delaying
the qualification exam to her fifth semester because she still did not
have an advisor.  She said, ``Professors are not helpful with students
gaining a background to be in a particular area of research.''  She
also reported that when she came to UIUC, she expected she could ``do
any research I wanted'' and didn't expect the research opportunities
to ``be so narrow.''  Another female participant said, ``Compared to
my male counterparts, I need more guidance [from my advisor] as to
what I'm doing.  I need more constructive criticism and more
constructive praise.`` A male student said,

\begin{center}
\emph{I wish someone had told me what I should be looking for in an
advisor, what I should be expected to do right away, what the
available research areas are, the delicate balance of taking classes
and doing research, and the funding problems in the department.}
\end{center}

He suggested that the students have a second-year review so that the
department can make sure they ``understand the process of research
from beginning to end.  The department thinks this is done by the
advisor.  This isn't necessarily true.''

Finding an advisor appeared to be especially difficult for women; a
difficulty which is compounded by a lack of female and minority
advisors in the department.  Female and other minority students are
unable to find someone like themselves or an advisor they think will
understand them.

\subsubsection{Mentoring}

Mentoring is one approach to facilitate a student's feeling of
membership in the Computer Science community.  Participants were
surveyed as to whether or not they had a mentor.  For those who did        
have a mentor, they reported on the sources of their mentoring
relationships.  For those who did not, they reported on the reasons
why.

\begin{figure}
\centerline{\psfig{figure=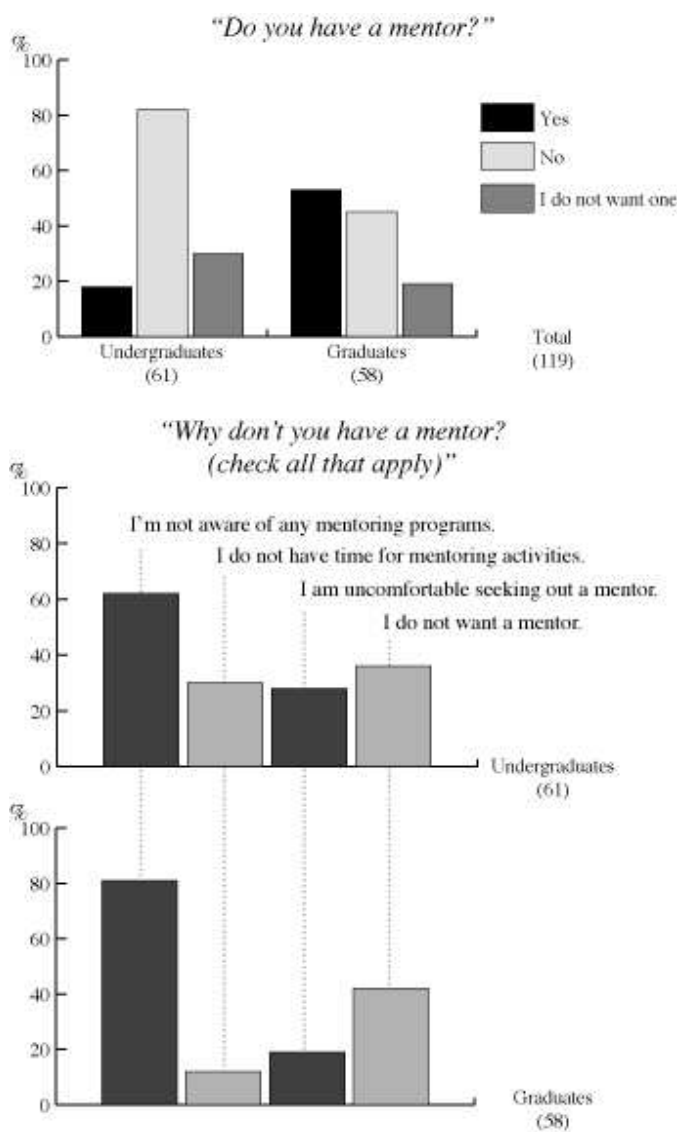,width=3.5in}}
\caption{Participants were asked if they currently had a mentor,
either within the department or from another source. Participants
without mentors also selected from possible reasons why they do not
have a mentor.  Many participants were unaware of any mentoring
programs within the Department of Computer Science.%
    \label{f:mentor}}
\end{figure}  

As summarized in Figure~\ref{f:mentor}, approximately 18\% of
undergraduate and 53\% of graduate students reported having a mentor.
For both groups, there was a gap between those participants who
reported not having a mentor and those who did not want a mentor,
suggesting that approximately 52\% of undergraduate participants and
26\% of graduate participants who currently do not have a mentor,
would like to have one.  Of those who reported not having a mentor,
approximately 62\% of undergraduates and 81\% of graduates report
being unaware of any mentoring programs available in the department.

\begin{figure}
\centerline{\psfig{figure=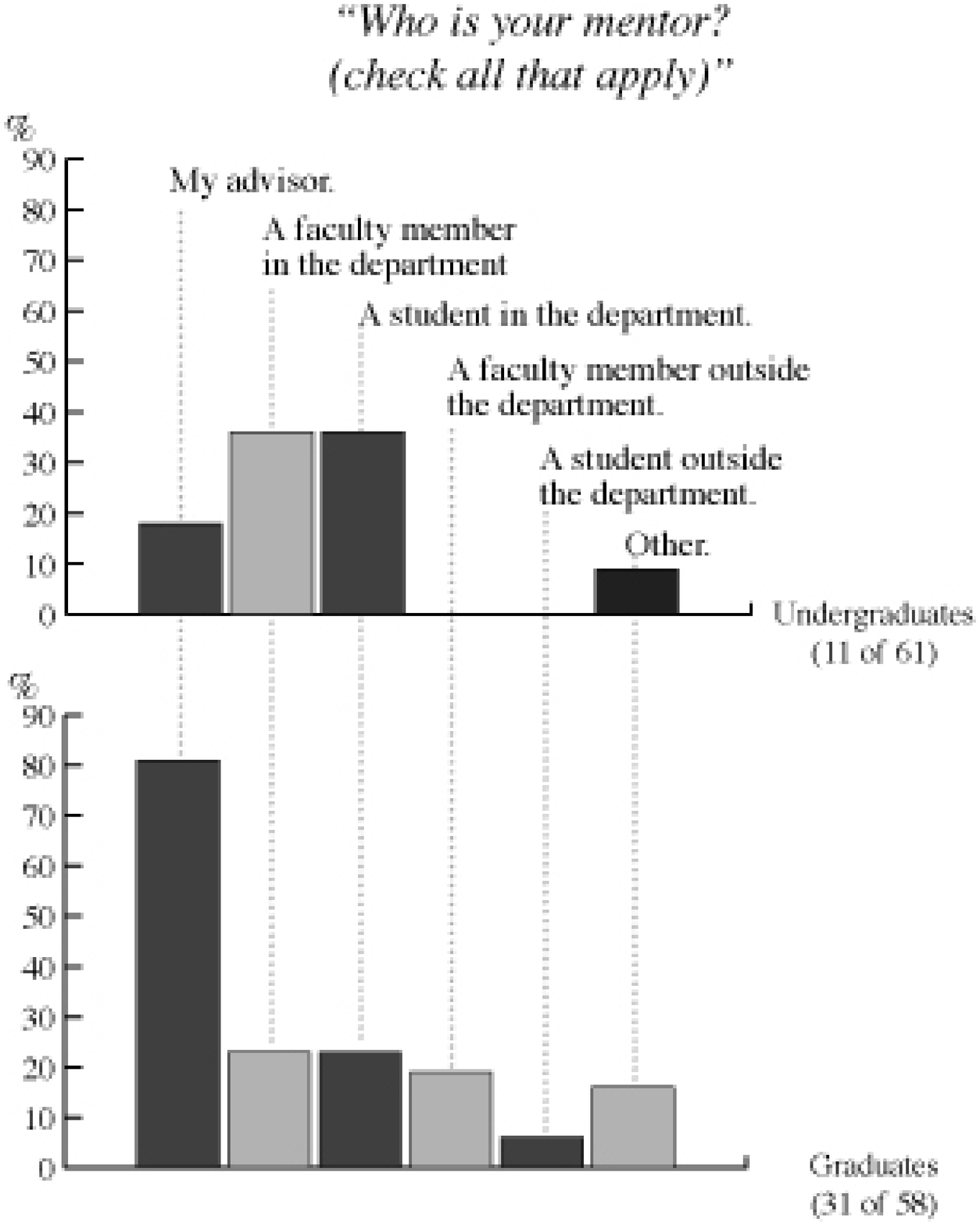,width=3.5in}}
\caption{Those participants who reported having a mentor were asked to
report on the source of their mentor. ``Other'' sources of graduate
mentors included those found on programs such as MentorNet, or past
employers.  For undergraduates, the single ``Other'' option selected
was a family member.%
    \label{f:mentor_source}}
\end{figure}  

During the interviews, we captured some of the attitudes of those
participants without mentors. One participant with an advisor said she
felt a mentor would be very valuable, but felt uncomfortable seeking
out a mentor.  She said, ``If I had to pick one tomorrow, I know who
it would be'' but couldn't bring herself to do it.  She said, ``There
is no formal system, which is unfortunate'' and suggested the value of
a formal system in which students could option for a mentor.  Another
participant without an advisor or mentor said, ``I did not expect the lack of
mentorship'' that she found at UIUC and felt that 

\begin{center}
\emph{The last thing the faculty cares about is chatting about my
problems.  No one would be interested in mentoring me. People here
just care about great research, not about mentoring.}
\end{center}

\newpage
Figure~\ref{f:mentor_source} summarizes the actual sources of mentors
for students in the department.  For the undergraduates, only 11 of
whom reported having a mentor, faculty members and students were
relied upon most for mentors.  WCS appeared to be a major source for
mentors for some participants, yet WCS is not the perfect support
solution for all of the participants.  WCS appears to be an effective
social support group for some subset of the female and even male
population of undergraduate students.  However, other students
mentioned having a more diverse set of interests than WCS could
address.  One participant said of her
disinterest in WCS,

\begin{center}
\emph{They have these coffee hours, but they don't have these ways to
break you in $\ldots$ for people to introduce you.  Most people in it
are all very $\ldots$ into video games and anime, and so I really don't
feel like I feel fit in.  When I did try to sophomore year, and I
found out that guys are in it, and they were the heads of whatever
$\ldots$ making projects group. Dude, I deal with enough guys already.
Where do the women come in?}
\end{center}

While this interviewee in particular expresses a need for a space
where she can escape the "guys" that she encounters on a daily basis,
many women involved in WCS appreciate the support of the men in the
department and welcome them into the group.  Striking a balance
between single and mixed gender groups is not easy, but research shows
that attempting to do so is necessary in order to support women as an
underrepresented group and to allow men to be active in creating a
climate which is more open to diversity \cite{NSF03}.

For graduate students, it was presumed a more natural mentor
relationship could be found in the advisor relationship.  In some
sense, this was the case; the majority of graduate students reported
their advisor as their mentor.  However, of the 51 graduate
participants who reported having an advisor, only 25, or 49\%,
reported that their advisor was their mentor.  This led to our
distinction between students who had an advisor relationship and an
``Advisor as Mentor'' or AAM relationship.

\begin{figure}[h]
\centerline{\psfig{figure=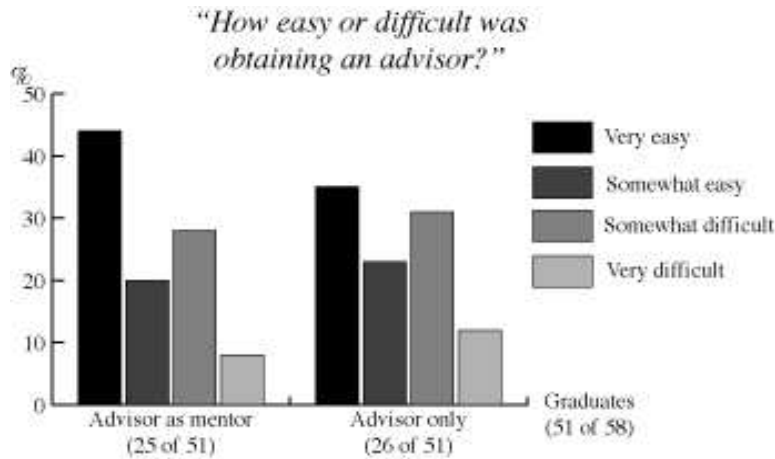,width=4in}}
\caption{Participants reported on the ease or difficulty with which
they obtained an advisor.  Those who reported that their advisor also
acted in a mentoring capacity reported a slightly easier time in
obtaining an advisor than those who did not.%
    \label{f:obtain}}
\end{figure}

To examine the impact on whether or not a graduate advisor was
considered a mentor by his or her student we looked at three areas.
First, we looked at how the ease of obtaining an advisor differed
between those who reported having an AAM and those who didn't.  As
reflected in Figure~\ref{f:obtain}, those with an AAM report a
slightly easier time of obtaining an advisor than those who do not.
Still, this isn't the entire picture.  In the interviews, multiple
participants discussed the hardships involved in obtaining an advisor
which were unrelated to the quality of the relationship with their
eventual advisor (see Figure~\ref{f:ease}).

Second, participants were asked to rate how supportive their advisor
is in helping them to become a successful computer scientist.
Participants rated their advisor as ``1'' for least supportive and
``5'' for most supportive.  Participants with an AAM gave an average
rating of 4.24 (standard deviation of .65).  Participants with only an
advisor relationship gave an average rating of 3.65 (standard
deviation of 1).

The reasons for which graduate students do not regard their advisors
as mentors vary.  In the interviews, many participants expressed a
reluctance to speak with their advisor regarding personal or even non
research-related issues.  One female participant with a male advisor
said, ``There are things I wish I could talk to my advisor about.'' She
also reported that she felt she would be able to talk to a female
advisor regarding issues she was unable to bring up with her male
advisor.  Other participants reported a disconnect between their
preconceived notions about a mentor and their own advisor.  
One male participant said of his advisor,\\

\emph{It's not like she's 60 with lots of experiences.  When I think
of a mentor, I think of some old guy who can pull strings for you to
get a job.} \\

Another male participant echoed this perception.  He didn't consider
his male, pre-tenured advisor as a mentor, but instead defined his
would-be mentor as, ``Someone who is on my side who has some
influence.''

\subsubsection{Discrimination}

Discrimination has the opposite effect of mentoring, in that it
alienates students and lessens their sense of membership in the
community.  In the pilot study, we inquired about participants
experiences with discrimination, using the following university
definition. 
\vspace{.05in}
\begin{center}
{\small The University of Illinois will not engage in discrimination
or harassment against any person because of race, color, religion,
sex, national origin, ancestry, age, marital status disability, sexual
orientation including gender identity, unfavorable discharge from the
military, or status as a protected veteran and will comply with all
federal and state nondiscrimination, equal opportunity, and
affirmative action laws, orders, and regulations.  This
nondiscrimination policy applies to admissions, employment, access to
and treatment in the University programs and activities.}
\end{center}
\vspace{.05in}

Among the graduate participants, all interviewed said that they had
not been discriminated against by faculty or students in the
department.  However, many of the participants proceeded to describe
situations which the interviewers felt constituted as some form of
discrimination.  For example, one woman reported that she had been
never discriminated against, but later told a story about how she'd
been told by a male student in computer science that ``women just
aren't as quick as men in computer science.''  She reported that this
caused her to be more self-conscious.  We can only guess at the
reasons why graduate students are more hesitant to describe these
kinds of events as discriminatory.  Because of these initial results,
we did not include discrimination questions in the second phase of the
graduate student portion of the study.

\begin{figure}[h]
\centerline{\psfig{figure=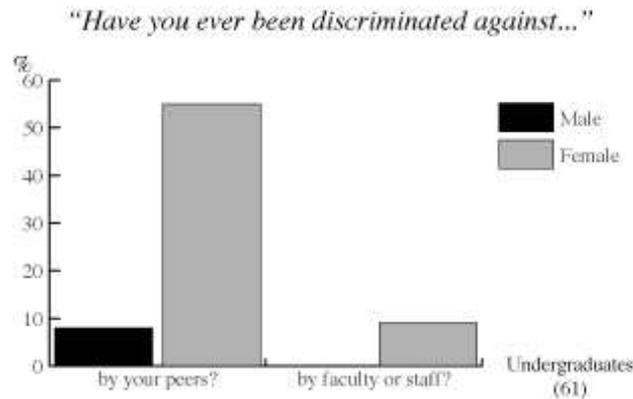,width=3.25in}}
\caption{Undergraduate participants were asked if they had ever been
discriminated against by their peers or faculty.  Among the men, none
reported any discrimination from faculty or staff, and 8\% reported
being discriminated against by their peers.  Among the women, 9\%
reported instances of discrimination by a professor or staff member,
and 55\% reported instances of discrimination by their peers. %
    \label{f:discr}}
\end{figure}  

All four undergraduates interviewed in the pilot study reported that
they had been discriminated against by their peers.  Three of the four
reported that they had been discriminated against only a few times,
and that they simply coped with it on their own.  One reported
frequent instances where she felt discriminated against, but said that
``many times, it was unintended or not hurtful to me, but it could
definitely be considered discrimination.''  She reported dealing with
it by speaking to both friends and departmental staff and faculty.

Given these initial results, we included a section on discrimination
in the second phase of the undergraduate portion of the study,
summarized in Figure~\ref{f:discr}.  Among the men, none reported any
discrimination from faculty or staff, and 8\% reported being
discriminated against by their peers.  Among the women, 9\% reported
instances of discrimination by a professor or staff member, and 55\%
reported instances of discrimination by their peers.

\begin{center}
\emph{It's funny how when I end up working with guys versus girls.
Like in [one course] last semester, there was a group of us $\ldots$
it was often me and these three guys, but they would always went off
on tangents about blah blah blah blah, and they would conveniently
finish the homework [later].  Most of them weren't organized, and it
always took us forever to get things done. I did feel that they didn't
necessarily defer to my opinion as much. If I give an answer then it's
not always valued.  My opinion should be valued as much as theirs.}
\end{center}

Regarding discrimination by faculty, one participant in the
second phase described her experience in one of her computer science
courses.

\begin{center}
\emph{My professor $\ldots$ all of his jokes are about like, 'oh guys
you can tell this to your girlfriend,' and then tells some random
joke.  That's funny, maybe the first time, but if you do it the whole
semester, there are girls in this class. It's not funny at the
end of the semester.}
\end{center}

While such an experience may not seem overtly discriminatory against
women, these kinds of small experiences can result in feelings of
alienation for women.

\subsubsection{Work and life balance}

\begin{figure}[h]
\centerline{\psfig{figure=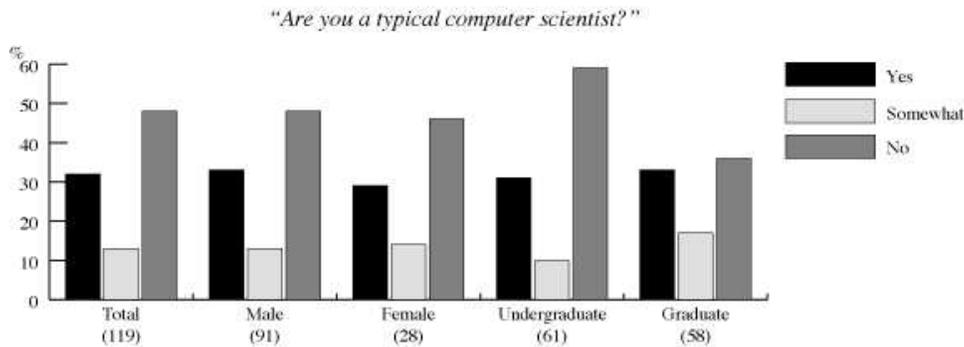,width=5in}}
\caption{Participants were asked if they considered themselves a
typical computer scientist.  The results were uniform across multiple
categories (gender, rank), with most students reporting that they did
not consider themselves as such.%
   \label{f:typical}}
\end{figure}

To begin, we asked participants if they felt they were a ``typical
computer scientist'' without offering any definition.  This allowed us
to gain insight into participants' own definitions of a ``typical
computer scientist'' and how they compared themselves to their own
definition.  Not many participants felt they were a typical computer
scientist, as seen in Figure~\ref{f:typical}.  Participants reasons
were incredibly varied, citing their particular research interest, the
number of computer languages they know, their race, their gender,
their looks and hygiene, their membership in a fraternity, their
interests in other topics, and other elements of their lifestyle as
reasons why they are not the typical computer scientist.  In fact, the
reasons for not being a typical computer scientist were so varied that
it was infeasible to categorize them.  Rather, we list a few here as
examples.

\begin{center}
\emph{No. I've noticed a lot of the people in the major don't look
like me and I'm not particularly enjoying the major at that.} \\[2ex]

\emph{No, because I do not enjoy the low level intricacies of computer
systems. I am more interested in building computer tools that have a
direct impact on human life and for whatever reason, that isn't
considered as pure computer science.} \\[2ex]

\emph{No.  I think we theoreticians are far too `math' to be typical.} \\[2ex]

\emph{No.  I'm more of a jack of all trades.  I like computer science,
but my life doesn't completely revolve around it.  I write; I read; I
make music; I cook.  Many of my fellow students appear one-sided.}
\end{center}

\begin{figure}[h]
\centerline{\psfig{figure=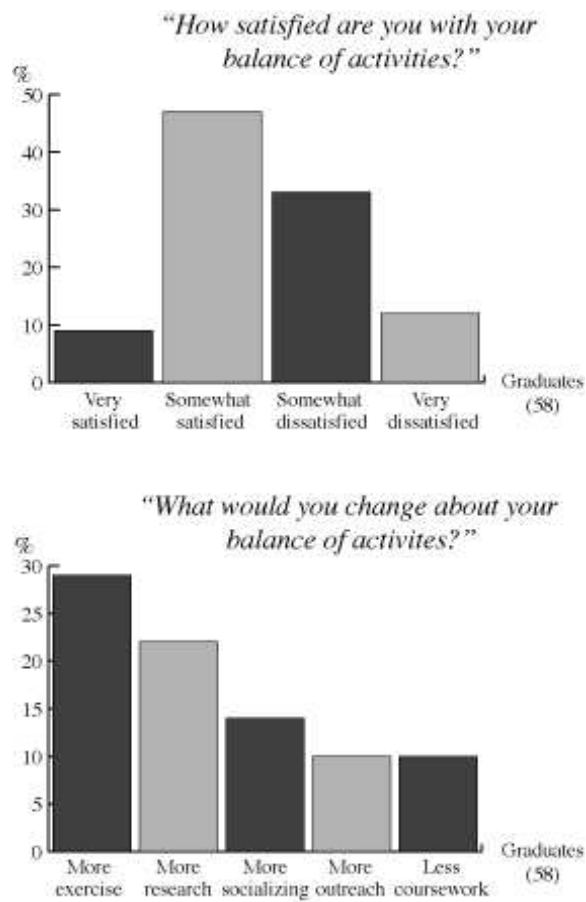,width=3in}}
\caption{Participants were asked how satisfied they are with their
balance of activities.  They were also given the opportunity to
express how they would improve the balance if they could in a short
answer format.  Despite being short-answer, participant responses were
easily categorized into the above themes, with most wishing for more
time for physical activity.
    \label{f:balance}}
\end{figure}  

Many of the graduates interviewed expressed a disinterest in leading
the lifestyle lead by the professors in the department.  One
participant, who was leaving the department this summer, noted a
divergence between the values of the department and his own personal
values.  He said, ``[Here] research is the only thing that matters.''
This is contrary to his own value which is, ``Do I get to spend time
with the people I care about?''  Another participant pointed out the
number of hours professors spent at work, saying,

\begin{center}
\emph{I think being a professor at a big state school is a very
difficult thing.  If I'm going to be a professor, I don't think I
could hack it here.  I don't think I could do what a lot of professors
here do.  I don't want to work 90 hours a week.}
\end{center}

Another participant echoed this statement with, ``I don't envy the
life of assistant professors at this school; working 100 hours a week
when you're 35.''

Given that the graduate school years overlap peak childbearing years,
we asked the graduate\footnote{Undergraduates were not surveyed about
family pressure or child-rearing based on the lack of results in the
pilot study.}  participants about their attitudes about parenting
while being a computer science graduate student at UIUC.  Of the
graduate participants, 9\% reported that they were a parent and spent
``10 to 20'' or ``30 to 40'' hours a week raising children.  Of the
participants, 35\% of the females and 10\% of the males said they felt
pressure to start a family.

In the interviews, the pressure to have a family was discussed.  One
participant, who was hopeful about someday starting a family, said
this of her pursuit of a Ph.D.

\newpage
\begin{center}
\emph{I shouldn't have to give up what I have here. I think that I can
have both. The problem is that most people don't think that way.  If
you are a female then a family should be your priority, not your
career $\ldots$ I think it's possible to have both. It's
difficult. It's not easy.}
\end{center}

Another graduate participant said of her reasons on deciding not to
have children,

\begin{center}
\emph{
I really don't see it happening, the way that I'm living now. I can
take care of a child at the high level in the sense that it won't
die, but basically you have to put the child in day care at 8 and take
it out at 6
$\dots$ Assume if I have a deadline, I don't even eat
properly if I have a deadline $\ldots$
Why would I do that to a child, basically. That's why I've given it up
altogether.}
\end{center}

To test whether the participants were aware of the the department's
existing family medical leave policy for graduate students, graduates
were asked, ``Does the Department of Computer Science at UIUC have a
family leave policy?'' Only 3\% of participants replied ``Yes'' while
91\% replied ``I don't know.''

The current policy states,

\begin{center}
   {\small Graduate students in need of a leave
   of absence due to medical or family emergencies, including the
   birth of a child, may request a one semester (or in rare cases an
   academic year) absence from their regular degree programs. Any time
   constraint on the degree requirements, such as the Master's degree
   time limit or time limits on Qualifying and Preliminary
   examinations, will be automatically extended by the length of the
   leave of absence. However, the Graduate College seven year clock
   for doctoral studies does not stop by a leave of absence. Any
   extension to this time limit must requested by petition to the
   Graduate College.}
\end{center}

When shown the policy, graduate participants were asked ``How adequate
do you consider this policy for your own family plans and needs,'' the
following answers were selected as follows:

\begin{itemize}
\item ``Very adequate.'' (16\%)
\item ``Somewhat adequate.'' (50\%)
\item ``Somewhat inadequate.'' (26\%)
\item ``Very inadequate.'' (7\%)
\end{itemize}

All of the parents reported that they felt the policy was ``Somewhat
adequate.''  Moreover, the parents interviewed felt that their advisor
was supportive of their role as a parent and that it was possible to
have a family while in graduate school. Still, some had suggestions
for improvements by the department; including a ``flexible schedule''
and ``1-2 month family leave.''

\section{Recommendations} \label{s:rec}

Given the data presented by our study, we provide a set of
recommendations in which the Department of Computer Science may
improve in the areas of recruitment, preparation, and retention.
These recommendations rely heavily on current practice elsewhere in
academia, seeking possible solutions for the particular areas of the
Department of Computer Science that need improvement.

\subsection{Recruitment}

    \noindent {\bf Provide more comprehensive information to
    prospective graduate students.}  The study highlighted the na\"{i}ve
    algorithm that graduate students used to select a graduate
    school.  This demonstrates a demand to make more information
    available regarding the choices that graduate students need to
    make when selecting a graduate school.  This information may
    better attract strong candidates who are informed about the
    department and graduate life at UIUC, and turn away students who
    might not succeed in the department's unique environment. One
    resource, the Survey on Doctoral Education and Career Preparation
    \cite{Golde01a}, offers a website \cite{Golde01b} summarizing the
    tough questions that prospective graduate students should ask of
    themselves and their prospective departments and programs of
    study.  Given that prospective students often visit the
    departmental website, linking {\bf cs.uiuc.edu} to third-party
    websites such as these help to provide prospective graduate
    students with the information they need to make the tough
    decisions about whether or not to attend graduate school and what
    graduate school to select. \\

   \noindent {\bf Facilitate more opportunities for outreach.} As
   noted in Section \ref{ss:org}, there are a number of existing
   outreach programs both within the department and the college.  The
   study showed that these are somewhat well-known by the
   undergraduates, but that the graduate students lack an awareness of
   these programs.  In discussions with faculty after initial
   publication of our  study, it seemed that the same was true
   of the faculty.  Even faculty are not consistently informed about
   outreach.  One simple solution to begin informing the department
   population of these efforts; include a slide or two about existing
   outreach efforts in the introductory presentations at the required
   seminar for all first-year graduate students or faculty meetings.

   The Department of Computer Science currently has several strong
   outreach programs, including ChicTech and CSGSO's prospective
   student weekend activities, and also participates in college-wide
   programs such as the G.A.M.E.S. camp.  However, there are two areas
   we feel that the existing outreach system could improve.  First,
   there is no central point of entry into these outreach programs.
   Individual programs perform their own recruitment via mass e-mail
   and word-of-mouth.  As mentioned in the survey, 18\% of
   participants currently participate in some kind of outreach, not
   necessarily within the department, for 1-5 hours per week.  These
   and other students who are interested in participating in outreach
   may simply be unaware of these recruitment efforts, resulting in an
   untapped though valuable resource.  A departmental-run umbrella
   organization would help students become aware of all outreach
   opportunities in the department, and might also allow the
   department to expand outreach to student organizations which
   currently do not participate.


   Second, there is no reward system for students participation in
   outreach.  To exemplify what we mean by reward system, two years
   ago the College of Engineering offered a course, Engr199, which was
   designed to allow students to help design a curriculum for Home
   High, a local school.  The course involved some reading and
   research into current practices, but also had a significant
   component in which students actually interacted with and taught the
   middle school students.  The Department of Computer Science also
   offered a similar course under the namesake ``CS2GO,'' whose recent
   offering yielded the MergeSort DanceTroop, a small band of
   graduates and undergraduates who perform a choreographed dance of
   the merge sort algorithm.  While schedules are invariably busy for
   both undergraduate and graduate students, the ability to receive
   course credit for outreach might allow interested students to
   participate in outreach activities.  This may be more applicable to
   undergraduates, but also to graduate students who are interested in
   gaining mentoring experience before becoming faculty
   themselves. Moreover, these opportunities allow students to
   participate in outreach who do not feel they fit in with ACM or
   WCS.

   In the undergraduate program, it is somewhat common for students to
   interact with companies or local organizations, which serves as an
   additional community outreach service.  For example, many students
   in CS427, \emph{Software Engineering I}, do software engineering
   projects based on corporate requests from local computer businesses
   like Motorola or Cisco.  The ChicTech program specifies that high
   school teams participating in the weekend competition should
   program something for a local non-profit organizations.  Projects
   identified and implemented by the high school students have
   included tracking software for their high school coach and creating
   a website and newsletter for a local women's shelter.  These
   opportunities to use computer science to benefit the real world,
   whether it is a local business or a non-profit organization, both
   accomplish community outreach and encourage students to consider
   the social impact of computer science.

   While these outreach efforts are all strong and successful at the
   undergraduate level, another concern is the lack of awareness of
   such programs by graduate students.  Recall the participants who
   said of outreach, ``I would if I were asked.''  or simply that,
   ``We don't do it.''  Given their perception that the department had
   few outreach activities, multiple graduate student participants
   offered great ideas for incorporating their research into outreach
   activities.  One participant offered an idea about workshops for
   people in industry, giving an example of ``rapid evaluation and
   prototyping.''  With this idea, the department could forge even
   more interactive relationships at the graduate level with local
   companies such as Motorola.  The message to industry would be,
   ``The things we are teaching you are the same thing we are teaching
   our students.  You should hire our students.'' Another said,

   \begin{center}
   \emph{In [a recent conference] there was this talk about how you
   can bring in the community in your $\ldots$ courses. So one professor was
   talking about how in her class she had students go out into the
   community and look for problems that were there and come up with a
   project that would address those problems.  Often in Computer
   Science, we pick problems that no one actually cares about, or no
   one even actually understands.  This is something how you can
   actually do community outreach.  You can go out to schools or Boys
   and Girls club and see what they are doing and how you can help
   them do certain things better. It's also applying what you are
   learning as a computer scientist.}
   \end{center}

   In the above a graduate student unknowingly describes the ChicTech
   program, and inserts an opportunity for herself to participate.
   Both undergraduate and graduate community outreach could be
   strengthened by encouraging new ideas and expanding existing
   programs.  Given that faculty are often unaware of business
   connections in the community, and graduates are often unaware of
   outreach opportunities, creating a centralized list of industry
   affiliates could help to foster such outreach, allowing interested
   students, research groups, or student groups like the CSGSO to
   organize these types of community ventures. \\

   \noindent {\bf Facilitate more interaction between students and
   faculty}.  In the final question of the survey, students were asked
   what they think could be improved in the department.  Of graduate
   participants, 31\% cited increasing student-professor interactions,
   and without prompting often cited the recently organized TGIF event
   as a good example.

    \begin{center}
      \emph{I think more informal gatherings between faculty and students
	would help the atmosphere. For example, the Friday lunches are a good
	idea. I would like to see more of those around the department.}
    \end{center}

    Another participant echoed this positive response to TGIF and went
    on to suggest more of these informal events,

    \begin{center}
    \emph{One suggestion is to promote some meetings between
    professors and students who have something in common. For example,
    meetings between a professor who has some outside interest that
    matches the interests of some students (e.g. basketball, cooking,
    kites etc).  I believe having something in common with a professor
    would make students more relaxed and facilitate a first contact
    and interaction.}
    \end{center}

    Other participants suggested organizing sports teams, bicycle
    trips, reflecting many participants' interest in increased
    physical activity.  Another participant suggested inviting faculty
    to the CSGSO FEs.

    Both ACM and WCS have regular meetings and seminars, and
    frequently invite faculty members to serve on panels and meet the
    students afterwards.  For example, WCS hosts a popular annual
    panel called ``Meet Your Professors.''  CSGSO has a few similar
    seminars, but they have generally been more limited and dependent
    on the particular CSGSO officer in charge of seminars.  Another
    suggestion to improve interaction between faculty and graduate
    students is to put senior graduate students in charge of
    coordinating talks and seminars, especially local seminars such as
    CS591, \emph{Advanced Seminar in CS}.  For larger seminars such as
    DCS and the Distinguished Lecture series, graduate students could
    be added to the committees that choose and arrange speakers.
    While graduate students cannot be expected to contact and arrange
    all such activities, they can be allowed to assist in planning and
    hosting.  These activities may help to eliminate some of the
    ``Us'' (the graduate students) and ``Them'' (the faculty)
    dichotomy that currently takes place at these seminars.  In
    addition, it would place senior graduate students in a more
    visible, prominent position, which in turn could help graduating
    PhDs to secure more contacts and better chances in the academic
    job market.

    Among undergraduates, 24\% suggested more faculty interaction and
    social activities to improve the department.  Several commented on
    how great !bang's events such as !Studybreak and !Casino were in
    terms of encouraging social activities and interactions with
    professors, and said more events of that type would be nice.  Two
    commented on how great the Powerlunch program was, but noted that
    most undergraduate students are too intimidated to ask a professor
    out to lunch themselves.  As a solution, one option is to have a
    weekly lunch sponsored by the department with limited seating.
    One or two professors per week would attend as ``hosts'', and
    students could sign up to attend.  In this way, more interaction
    with professors is accomplished, but in a less intimidating
    setting and where students could meet professors whom they might
    never have interacted with before.

    As another option, it might be beneficial to encourage
    undergraduates to talk to their advisors and professors during
    office hours more frequently.  However, many undergraduates
    misunderstand the purpose of the advisor, assuming that faculty
    advisors are there to assist with course planning.  Better
    informing them of their advisor's purpose, perhaps during CS100 or
    other 100- and 200-level required courses, may encourage students
    to regard their advisors as mentors with regards to career
    guidance and other professional advice.  Resolving this
    misunderstanding might also help faculty advisors serve their
    students more effectively.

\subsection{Preparation} 

    \noindent {\bf Improve quality of teaching}.  While it is
    understood by the participants that the department's predominant
    emphasis is research, the quality of teaching was still raised by
    both undergraduates and graduates.  For the undergraduates,
    improving the quality of teaching means improving the quality of
    teaching assistants and to fostering more collaborative course work.

    Before discussing how to improve teaching assistants, we summarize
    the current system for training and assigning TAs in the
    department.  Currently, teaching assistants for the department
    receive the same two-day training course that all campus-wide
    teaching assistants receive.  International teaching assistants
    receive an additional two days of training to help them cope with
    the gap between their own culture and American culture.  The
    Department of Computer Science assigns teaching assistants to
    courses with the requirement that they've taken the course before.
    However, because of lack of funding, or lack of teaching
    assistants with a certain expertise, graduate students are
    sometimes assigned to courses they've never taken.  Moreover,
    professors are decoupled from the assignment process, with their
    assistants sometimes lacking the expertise they expect for the
    particular course as its taught at UIUC.  On the other hand, some
    graduate students are particularly gifted at teaching,
    consistently lack funding for a research assistantship, or
    both. This results in career teaching assistants, graduate
    students who spend two years or more TA-ing courses.  One
    participant was a career TA, and reported leaving the department
    without completing a PhD this summer.

    That said, the current system for training and assigning teaching
    assistants appears that it could be improved.  First, given what
    little training is available to teaching assistants, it may be
    worthwhile to provide more resources to them.  Currently, the
    university is host to CTEN, the College Teaching Effectiveness
    Network.  It sponsors events throughout the year, including talks
    on increasing student motivation, and grading assignments.  It may
    be worthwhile to for the department to better utilize this
    existing resource by advertising CTEN to its TAs who might opt for
    additional training.  A step further would be for the department
    to offer more specialized training for its teaching assistants.
    For example, the University of California at Berkeley offers its
    own course, CS301, called \emph{Teaching Techniques for Computer
    Science} which discusses techniques for effective teaching.

    In terms of assigning teaching assistants, professors could be
    given an opportunity to participate, interviewing the available
    teaching assistants to determine if a student's expertise is a fit
    for the course.  To increase the available expertise in the pool
    of teaching assistants, it may be worthwhile to require a year of
    teaching service of all graduate students, thereby gaining the
    expertise of third-year students who've never taught while at the
    same time avoiding career teaching assistants.  Incoming
    first-year graduates do not always have the expertise to TA
    courses such as CS423, \emph{Operating System Design}, or CS473,
    \emph{Algorithms}.  Distributing the one year service requirement
    across the entire graduate program could help obtain more
    expertise.  Finally, given that approximately 40\% of
    Ph.D. candidates reported an interest in a career at four-year
    university, requiring a year of teaching from graduate students
    may better prepare them for their future career choice.

    As previously pointed out, undergraduate participants requested
    more collaborative course work, whether through discussion
    sections or group assignments.  Several students specifically
    commented on the new course organizations for the CS\emph{x}73
    sequence.  Many liked the new problem sessions in CS273,
    \emph{Intro to Theory of Computation}, and CS473U, the
    undergraduate section of \emph{Algorithms}. The discussion
    sections in CS173, \emph{Discrete Structures} were also mentioned
    favorably.  These interactive learning environments appear to help
    students who learn more effectively outside of traditional
    lecture, or for students who might be seeing the topics of CS173
    for the very first time.  In addition, they provide more social
    interaction among undergraduates and between graduate students and
    undergraduates.

    For both undergraduates and graduates, improving the quality of
    teaching means improving teaching styles.  One participant
    said,

     \begin{center}
       \emph{I would like the university to help teach professors who are great
	 minds/researchers how to teach and interact with students better.  I
	 understand that UIUC is a great research school, and I know the
	 importance of this, but sometimes the people who are best at research
	 are terrible teachers.}
     \end{center}

    That said, one recommendation is to offer a computer science
    pedagogy course to the department's incoming assistant professors
    who might lack teaching experience.  Currently, the College of
    Engineering offers such a course, but the approaches to teaching
    computer science aren't necessarily similar to those for teaching
    engineering courses.  Such a course would be taught by tenured
    professors of the Department of Computer Science, optional for
    tenured professors, but required for assistant professors.
 
    Another solution is to adopt the ``Assistant Professor of the
    Practice'' position \cite{Fogg04} that is seen at universities
    like Duke and Carnegie Mellon University \cite{CMU01}.  With more
    permanent lecturers who are more focused on teaching than
    research, students may gain more quality teaching for their
    foundational 100 and 200 level courses.  Such an improvement
    coincides with the multiple comments from participants who
    described their excellent experiences with teaching and mentoring
    as provided by current lecturers of such courses in the
    department. \\

    \noindent {\bf Provide more flexibility in core requirements}. The
    attitudes regarding the core curriculum requirements at the
    university are very disparate among the department and and
    faculty.  The department expects graduate students to achieve
    ``A'' grades in many of these courses, while it is understood by
    the participants that many of the faculty feels that course work is
    very unimportant compared to research.


    Currently, the department allows graduate students to take their
    qualification exam without meeting one of their core requirements,
    allowing them to postpone the course until after the exam.  This
    could be extended further, requiring that the core requirements
    for PhDs be met over three years, rather than two.  This way,
    students may take the courses important for their qualification
    exam, gain knowledge in foundational areas, and better balance
    course work and research. \\

    \noindent {\bf Increase early research opportunities}. Given the
    department's interest in reducing the time to degree, we recommend
    offering more opportunities to ease the transition from student to
    researcher, both for undergraduates and graduates, by making
    research opportunities available earlier in their programs.  In
    this regard, the department has already offered some opportunities
    which were met with positive response.  These include career
    panels of faculty during the DCS seminar as well as stand-alone
    talks such as Professor Sha's ``How to do Research.''

    More opportunities, however, could ease this transition further
    and reduce the current average time to PhD.  
A good example of
    such an opportunity is found at The University of Illinois
    Department of Mathematics.  They had a five-year NSF-funded
    program, starting in June 2000, called the Vertical Integration of
    Research and Education in the Mathematical Sciences (VIGRE)
    \cite{Vigre00} program. VIGRE provided for a number of work groups
    which fostered interaction among undergraduates, graduates,
    postdoc and faculty.  Among them were the Across Level Peers (ALP)
    group which included undergraduates, graduates, and faculty to
    discuss a variety of topics, including career plans, and foster
    mentoring relationships.  Also offered was the Research Experience
    for Graduate Students (REG) work groups, which were organized by a
    faculty and focused on a single research area.  Faculty, postdocs,
    and graduate students, including first-year students, would meet
    and take turns presenting problems.  As the seminar progressed,
    the students and faculty would break into smaller groups to
    conduct original research.

    At least one VIGRE group, the Combinatorics REG, continues to meet
    every summer, fall, and spring semester, despite the funded
    program's completion in 2005.  Participation has become one
    successful way to for first-year students in both math and
    computer science to enter into research in combinatorics as well
    as find an advisor.  Notably, many students in theoretical
    computer science have also participated in the summer sessions to
    broaden their research experiences in mathematics.  Almost all
    participants are able to produce a publication with others before
    the conclusion of the seminar each session. 

    The NSF has similar grants for Computer Science, including the
    ``Broadening Participation in Computing'' program.  This kind of
    program at UIUC could not only help graduate students make the
    transition to researcher, but it would also give undergraduate
    students an opportunity to learn more about research career
    opportunities.  Given that only 24\% of the male and 9\% the
    female participants reported that they would consider graduate
    school, this may help to increase the number of University of
    Illinois undergraduates who continue to graduate school.

    For the 41\% of female graduate participants who had a ``somewhat
    difficult time'' finding an advisor, we recommend programs like
    the one at the University of Washington \cite{Hand05}.  The
    Faculty and Graduate Mentorship Program promotes mentoring
    relationships between female graduate students and faculty in the
    science, technology, engineering, and mathematics fields.  These
    kinds of specialized mentoring programs between faculty and
    underrepresented minorities | like women and African Americans |
    could help to ease their existing frustrations and challenges in
    finding an advisor.

\subsection{Retention} 

    \noindent {\bf Create multiple and diverse mentoring programs.}  The
    study uncovered a notable disparity between the participants who
    did not want a mentor and the participants who did not have a
    mentor, suggesting that there is a large population of students
    who are interested in gaining the benefits of a mentoring
    relationship.  For undergraduate students, aforementioned programs
    such as the VIGRE All Level Peers groups are designed to foster
    mentoring relationships between undergraduates and graduates and
    undergraduates and faculty. 

    For graduate students, there are currently few mentoring
    opportunities, the primary of which relies on a student's single
    advisor, yet only 49\% of graduate students with an advisor
    regarded their advisor as their mentor.  Realistically, no single
    advisor can be the perfect mentor, so we suggest providing
    multiple and diverse options for graduate student mentors.  For
    example, the University of Southern California's Department of
    Mathematics assigns mentoring triplets at the beginning of the
    academic year.  These triplets consist of a first-year graduate
    student, an advanced graduate student, and a faculty member.  It
    is the faculty member's responsibility to schedule approximately
    monthly meetings with the triplet for informal discussions.  Given
    the recent success of the TGIF program at UIUC, these further
    informal gatherings may help to foster an improved intellectual
    community in the department.  The triplets could also consist of
    freshmen, upperclassmen, and advanced graduates.  Yet another
    option may be to simply provide incoming freshmen or first-year
    graduate students with a list of upperclassmen and advance
    graduate students who are interested in being mentors, so that
    first-year graduates can option for a mentor if they wish.  These
    kinds of programs help mitigate the burden of mentoring on the
    faculty, improve interaction among undergraduates, first-year
    graduates, and advanced graduates, and may also help to change the
    perception that a mentor has to be, as one participant put it, ``a
    wizened old man.'' \\
 
    \noindent {\bf Provide an adequate family leave policy.}  Though
    the peak child birth years overlap graduate school years, there
    are currently few family leave policies nationwide for graduate
    students. This is notable for the department's female students,
    35\% of whom feel pressure to have a family.  Noting other
    studies, including \cite{Mason04}, which illuminate the need for
    such policies in order to promote greater female participation in
    academia, top schools are starting to implement assistance for
    students who choose to have children in graduate school. 

    The current leader, MIT, has a Childbirth Accommodation Policy. This
    policy, administered by the Graduate Students Office, allows up to
    eight weeks of Childbirth Accommodation. Students who are research
    and teaching assistants paid by the university continue to receive
    their stipend during this time. Teaching assistants are permitted
    to consider limited duties. MIT doesn't treat this accommodation
    as a ``leave'' as that could negatively affect some visa status.

    Other universities, such as University of California at Berkeley
    and Stanford University, have similar policies.  Student parents
    at Berkeley can request ``part-time status'' meaning that the
    course requirements are fewer, preliminary and qualification exam
    clocks are slowed, but full-time benefits are still provided.
    Pregnant women may also request part-time status.  Moreover, for
    four weeks before and 6 weeks after the birth, research advisors
    are instructed to expect that the student will be less productive
    than usual, and that the mother may request a temporary suspension
    of her research or instructor assistantship.  

    Graduate students at Stanford also have a Childbirth Policy in
    which female graduate students are eligible for two-quarter
    Accommodation Period in which examinations and other academic
    requirements may be postponed.  During this time, they are also
    eligible for full-time enrollment and keep their health insurance
    and university housing.  Given the university's population of
    5500, one-third of which are women, Stanford estimates that 30 of
    their female graduate students per year will participate,
    resulting in an annual cost of ``less than \$100,000''
    \cite{Cap06}.

    The existing family leave policy in the Department of Computer
    Science slows the clock on requirements for the qualification and
    preliminary exams.  Thus, a student could potentially be absent
    for a semester while recovering from her birth without suffering
    any penalty for not taking the preliminary exam on time.  However,
    current parents of the department only rated the current policy as
    ``somewhat adequate.''  To achieve a more adequate policy, we
    examine how the policy differs from that of MIT, Stanford, and
    Berkeley.  First, it currently relies heavily on a cooperative
    relationship between an advisor and student.  It does not have the
    same level of protection for students.  A student is not
    guaranteed the ability to return to her research or teaching
    assistantship after being absent for a period of time.  Moreover,
    an existing university-wide policy only provides graduate students
    with two weeks of leave for emergency or other health reasons.  A
    student who chooses to be absent for a longer period of time, like
    the minimal six weeks necessary to recover from a cesearean
    section, suffers from not having access to health insurance, or
    partial or full pay.  The Federal Family and Medical Leave Act of
    1993, as observed by the University of Illinois, provides that its
    eligible full-time employees, ``are eligible for up to 12
    workweeks of paid and/or unpaid leave'' \cite{UIUC93}.

\section{Conclusion}





    Overall, the study saw two major themes.  First the participants
    described an overall perception of the department placing
    significant emphasis on research pursuits.  These pursuits bring
    funding to the department and inform the broader computing
    community of advancements in computer science.  Yet, our
    participants also reported perceiving a negative impact as a
    result of this emphasis.  We spoke with participants who felt that
    professors sacrificed quality teaching for research and
    participants who felt that no professor would be interested in
    mentoring them.  We surveyed graduate participants who disdained
    the core course work for taking time away from
    research. Participants turned away from careers at top research
    institutions because of an unwillingness to pursue the perceived
    ``90 hour'' work week exemplified by many professors in the
    department. Still other participants reported leaving the program
    without completing their degree.

    As a highly ranked department in a large research university, it
    is understood that the Department of Computer Science at UIUC
    places a primary emphasis on research.  However, as \cite{Serow}
    and \cite{Baez} show, and our study reveals, this may be at the
    detriment to other roles | teaching and service | which are also
    important to a high quality academic environment.  Given the
    results of our study, we argue not that research should be placed
    secondary to these two other roles, but rather that a greater
    balance of this triumvirate is necessary to excel as a department.

    In a system which places a much higher importance on research,
    faculty who wish to teach well go unrewarded by the tenure system
    and even suffer from the public perception that their teaching
    activities detract too much from a focus on research.  Likewise,
    students may receive poor teaching even at schools which are
    highly ranked because research takes center stage.  For graduate
    students, mixed messages are received when the need to do research
    is combined with core requirements that do not necessarily advance
    the research goal.  In this case, pressure to obtain a certain
    grade in a course unrelated to one's research area surfaces as
    concerns that the courses are taking away from research and in
    advice from advisors that the core requirements "don't really
    matter."

    With regards to service, the push to do research may also
    discourage both faculty and graduate students from acting as
    mentors and performing outreach and recruitment activities as well
    well as make it difficult to find a healthy work and life balance.
    Similarly, underrepresented students and faculty tend to be
    targeted to be active in particular service activities, such as
    female faculty or graduate students receiving requests to act as
    mentors for female students or sit on gender-related committees,
    which do not aid in their promotion or in receiving their graduate
    degrees \cite{Baez}.  These additional expectations further
    compound the difficulty of being a minority in the department.

    Admittedly, our appeal for greater balance among the three roles
    of research, teaching, and service may seem like too great a
    challenge with too much a cost for the research pursuit.  Yet this
    triumvirate, when balanced, can lead to a positive environment in
    which each role informs the other.  For example, the encouragement
    to serve as mentors and perform other service activities
    frequently provides an opportunity to expand research
    opportunities.  Excellent teaching provides students with a
    stronger background with which they may do great research.
    Interaction within the department yields new graduate students and
    faculty with which to collaborate.  The roles do not need to be in
    competition with each other, but rather can be cultivated to
    complement each other. As we have seen while talking to both
    faculty and students, some pieces of this balance puzzle are
    already thriving within the department.  Outreach programs like
    ChicTech, CS2GO, and the G.A.M.E.S. camp allow undergraduates and
    graduates to use their computing expertise to impact K-12
    education.  Courses such as CS427, \emph{Software Engineering I},
    make connections to industry allowing students to network and
    demonstrate their excellence to outside members of the computing
    community.  Teaching improvements seen in the \emph{x}73 sequence
    foster more interaction among students, paving the way for more
    research collaboration in the future.

    This brings us to our second theme; simply the lack of
    communication among various entities within the department.  This
    lack of communication has a significant effect on the perception
    of participants that research is of primary value.  There are
    multiple outreach programs available and improvements in teaching
    are underway as seen in the CS\emph{x}73 sequence, yet many
    graduates and undergraduates are not exposed to these phenomena.
    A graduate participant complained that the department doesn't do
    any outreach, yet we just saw the conclusion of another successful
    G.A.M.E.S. camp on the local news.  Multiple participants
    complained about certain faculty needing to ``sell'' the students
    more, yet many other faculty speak highly of the graduate
    students, particularly with industry representatives and alumni.
    Undergraduates complained of a lack of interaction with faculty
    while faculty complained that students never attend office hours
    or come to speak with them.

    While this dichotomy was especially frustrating during the course
    of the study, it was also hopeful to observe how many efforts are
    currently underway.  With activities ranging from the TGIF lunch
    to !Casino, many of the faculty in the department are already
    involved in improving the student experience at an unprecedented
    rate.  While pieces of the balance do exist, more are needed.  It
    is not simply advertising current activities that is needed,
    though more diverse footage on the video wall would be welcome by
    everyone.  What is also needed is increased communication between
    students and faculty regarding everyone's needs and expectations.
    As one participant said, it is making sure that students
    ``understand the process of research from beginning to end.''


    The authors of this work were primarily motivated to conduct this
    study with the belief that its results and recommendations would
    serve as a list of best approaches to continue this already
    existing effort.  It is our hope that this document and our
    recommendations will open up channels of communication so that
    students and faculty can work together to better the department,
    and in turn, improve the computing culture for everyone.

\section{Acknowledgments}   

    Many people assisted with the creation of this study.  In
    particular, we would like to thank Marc Snir and the Department of
    Computer Science for supporting and funding our study.  Our thanks
    to Margaret Fleck for her evaluation of our report and helping to
    make it a Technical Report.  Our thanks also goes to the National
    Center for Supercomputing Application for their additional funding
    and to the Women in Computer Science who sponsored and supported
    our efforts.  Jeff Erickson, Deb Israel, Sam Kamin, Robin Kravets,
    Steve Lavalle, Klara Nahrstedt, Lenny Pitt of University of
    Illinois and Jennifer Croissant of the University of Arizona all
    provided helpful feedback.  Most of all, we would like to thank
    all of our study participants, who thoughtfully articulated their
    own experiences in the Department of Computer Science.

\bibliographystyle{acmtrans}
\bibliography{result}

\end{document}